\documentclass{article}

\usepackage{graphicx}
\usepackage{amsthm}
\usepackage[centertags]{amsmath}
\usepackage{amsgen}
\usepackage{amscd}
\usepackage{amsbsy}
\usepackage{amsopn}
\usepackage{amstext}
\usepackage{amsxtra}
\usepackage{amsfonts}
\usepackage{epsfig}
\usepackage[a4paper]{geometry} 

\usepackage[english]{babel}

\newcommand{\Te}{T_{eff}}

\newcommand{\bear}{\begin{array}}
\newcommand{\ear}{\end{array}}

\newcommand{\beq}{\begin{eqnarray}}
\newcommand{\eeq}{\end{eqnarray}}
\newcommand{\beqa}{\begin{eqnarray}}
\newcommand{\eeqa}{\end{eqnarray}}
\newcommand{\no}{\nonumber}

\def\OMIT#1{{}}
\newcommand{\lsim}{\mathrel{\rlap{\lower4pt\hbox{\hskip1pt$\sim$}}
    \raise1pt\hbox{$<$}}}         
\newcommand{\gsim}{\mathrel{\rlap{\lower4pt\hbox{\hskip1pt$\sim$}}
    \raise1pt\hbox{$>$}}}         

\begin{document}

\vskip1.5cm
\begin{center}
  {\Large \bf DAMA vs. the annually modulated muon background}\\
\end{center}
\vskip0.2cm

\begin{center}
{\bf Kfir Blum}

\end{center}
\vskip 8pt

\begin{center}
{\it Weizmann Institute of Science, Rehovot 76100, Israel\\ and\\
Institute for Advanced Study, Princeton, NJ 08540, USA} \vspace*{0.3cm}

{\tt  kblum@ias.edu}
\end{center}

\vglue 0.3truecm

\begin{abstract}
We compare the DAMA signal to the muon flux underground, which is annually modulated due to temperature variations in the stratosphere. We show that the muon flux at LNGS and the DAMA signal are tightly correlated. Different mechanisms were proposed in the literature by which muon-induced events may dominate the signal region in DAMA. We discuss simple statistical constraints on such mechanisms and show that the DAMA collaboration can falsify the muon hypothesis, if it is wrong, by reporting their annual baseline count rates.
\end{abstract}

\section{Introduction}

The DAMA/LIBRA experiment~\cite{Bernabei:2008yh} is a radiopure NaI(Tl) scintillation detector located at the Gran Sasso National Laboratory (LNGS), searching for the annual modulation signature of dark matter particles. With integrated exposure larger than 1 ton$\times$year and covering 13 annual cycles, DAMA/LIBRA and the former DAMA/NaI results reveal a clear annual modulation in the count rate of scintillation events~\cite{Bernabei:2008yi,Bernabei:2010mq}. The modulation period, phase and amplitude and the fact that modulation is only reported to exist in single-hit events, where only one NaI(Tl) crystal out of the 25-crystal array is triggered, are all consistent with expectations from direct detection of dark matter~\cite{Drukier:1986tm}. 

A thorough investigation of possible background processes which may explain the DAMA result is clearly of utmost priority~\cite{Kudryavtsev:2009gd,Ralston:2010bd,Nygren:2011xu,Schnee:2011ef}. The DAMA collaboration argues that no background was found which satisfies all the features attributed to the modulation signal~\cite{Bernabei:2008yi}, including the time dependence, amplitude and event distribution in the detector array. Despite of those arguments, one particular source of background, common to all underground low-noise experiments, merits very careful consideration when one tries to search for an annual modulation. This background is the flux of penetrating underground muons~\cite{Ralston:2010bd,Nygren:2011xu}. 

Annual variations in the muon flux underground have been measured by many experiments~\cite{lvd,Ambrosio:1997tc,Ambrosio:2002db,:2009zzh,Tilav:2010hj}. Modulation was also reported for neutrons~\cite{icarus} which are produced, among other mechanisms, through muon interactions with rock. 
In the northern hemisphere the muon rate peaks around June-July, close to the DAMA phase. The coincidence of the modulation in phase and amplitude motivated Refs.~\cite{Ralston:2010bd,Nygren:2011xu} to consider underground muons and/or neutrons as a possible explanation to the DAMA signal.

In this paper we investigate further the hypothesis that muons may explain the DAMA anomaly and propose simple methods to test it. In Sec.~\ref{sec:test} we extend the discussion in~\cite{Ralston:2010bd,Nygren:2011xu} by comparing the DAMA signal to the muon background directly. We point out that a muon explanation for DAMA cannot rely on muons directly traversing the detector material, but instead should involve muon interactions in a larger volume surrounding the detector. We study the time behavior of the muon rate and show that the DAMA collaboration can rule out the muon hypothesis, if it is wrong, using only existing data. This can be done exploiting the long term variability of the baseline muon flux, which is not expected from a dark matter signal. We conclude in Sec.~\ref{s:disc}. In App.~\ref{a1} we recapitulate the physics of the muon time variability.

\section{Testing the muon background hypothesis}\label{sec:test}

We begin by describing the DAMA signal. 
DAMA applies a software threshold of 2 KeV, in order to avoid the tail of PMT noise peaking at lower energies. Here, for concreteness, we focus on the energy interval 2-4 KeV. 
In this interval, using Refs.~\cite{Bernabei:2000qi,Bernabei:2008yi} we obtain the baseline count rates averaged over four DAMA/LIBRA and two DAMA/NaI annual cycles, respectively:
\beq\label{eq:damas}\label{s} \bar s&=&1.15\;{\rm cpd/kg/KeV\;\;\;\;\left(DAMA/LIBRA\,1-4\right)},\no\\
\bar s&=&1.45\;{\rm cpd/kg/KeV\;\;\;\;\left(DAMA/NaI\,3-4\right)}.\eeq

While Eq.~(\ref{eq:damas}) gives the baseline rate averaged over a few cycles, it is important to note that DAMA defines the ``baseline count rate" separately for each annual cycle~\footnote{We thank Rita Bernabei for clarification on this point.}. 
The residual count rate $\delta s$ in DAMA/NaI and DAMA/LIBRA was given in~\cite{Bernabei:2008yi} and~\cite{Bernabei:2010mq}. 
The residual count rate in each annual cycle is obtained by subtracting the  baseline rate of that cycle from the data.

\begin{figure}[h]\begin{center}
\includegraphics[width=15cm]{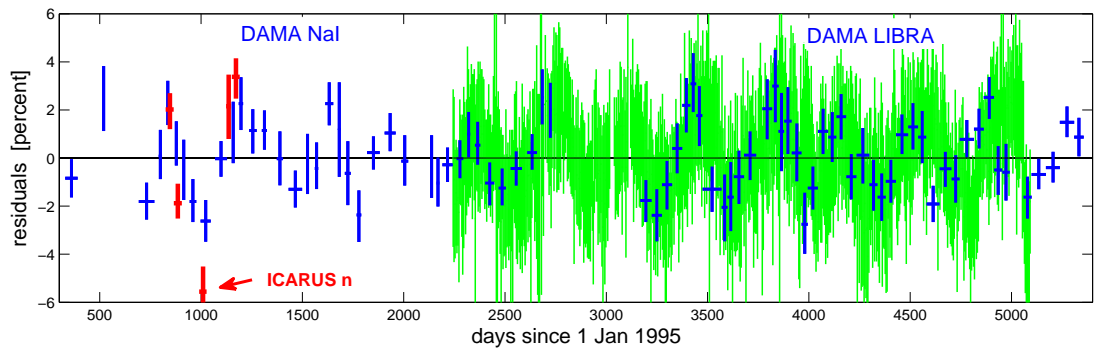}
\caption{DAMA residuals (blue~\cite{Bernabei:2008yi,Bernabei:2010mq}) and LVD muon intensity residuals (green~\cite{lvd}) in percent from the respective baselines. ICARUS neutron measurements during 1997-1998 are added (red~\cite{icarus}).}
\label{fig:1}
\end{center}
\end{figure}

Next, we require information on the muon flux. Measurements of the daily muon intensity at LNGS were carried out by the LVD experiment~\cite{lvd}. In Fig.~\ref{fig:1} we plot the DAMA/LIBRA and DAMA/NaI residuals (blue), in the form of percent modulation relative to the mean baseline count rates of Eq.~(\ref{eq:damas}), on top of the LVD muon intensity modulation (green)~\footnote{The publicly available LVD data~\cite{lvd} is provided in a dense graphical form. In converting the muon data to digital form, we end up with a reduced data set sampled roughly on a five-day basis. Our procedure carries the LVD daily error bars to the digitized five-day data, losing a potential factor of two improvement in accuracy.}. In presenting the muon  modulation we measure the residuals with respect to the baseline muon rates at each annual cycle separately, starting each cycle on Sep 9. Fig.~\ref{fig:1} makes clear that the muon background should be regarded with care. 

The DAMA collaboration considered the possibility that their signal is associated with muons~\cite{Bernabei:2009du} or with the fast neutron flux that muons induce~\cite{Bernabei:2003za,Bernabei:2008yi}. These possibilities were discarded based on three types of arguments:
\begin{enumerate}
\item Intensity: Considering fast neutrons, a rough estimate of the flux was made, indicating that the fast neutron flux is of order $10^{-3}$ cpd/kg/KeV, three orders of magnitude smaller than the DAMA count rate. This small flux was discarded as harmless.
\item Background rejection: Muon-induced processes (including fast neutrons) were argued to fail a number of background rejection criteria. The relevant criteria include the presence of multiple-hit events and deposition of energy above the signal region.
\item Modulation phase: The authors of~\cite{Bernabei:2009du} entertained the possibility that muon-induced events evade somehow the background rejection and intensity arguments, and considered the muon flux modulation of LVD~\cite{lvd}. Based on sinusoidal fits to both observables -- the DAMA residuals and the muon rate -- Ref.~\cite{Bernabei:2009du} concluded that the inferred phases of the two effects are separated by more than $5\sigma$.
\end{enumerate}
All of the three arguments above were challenged in~\cite{Ralston:2010bd} and~\cite{Nygren:2011xu}.

First, two mechanisms were discussed wherein muon-induced interactions in and around the DAMA apparatus could evade both the intensity and the background rejection criteria. Ref.~\cite{Ralston:2010bd} discusses in detail the neutron background, arguing that it is possible for neutrons to deposit energy in the DAMA signal region without impacting the higher energy region, and showing that the yield of  muon-induced neutrons may have been underestimated by orders of magnitude. The DAMA signal  could be associated to $\sim$ 3 KeV Auger-L electrons and X-rays accompanying the decay of $^{128}$I (lifetime $\sim$ 36 minutes), formed from $^{127}$I by neutron capture. Ref.~\cite{Nygren:2011xu} points out to the possibility of delayed phosphorescence in NaI(Tl), which could be initiated by muon-induced energy deposition inside the detector but trigger long after the actual muon passage, evading coincidence veto. We have nothing to add to the discussions of background rejection in~\cite{Ralston:2010bd,Nygren:2011xu}, and we refer the reader to these references for more  details. 
We do have further insight regarding the intensity argument, which we consider in Sec.~\ref{ssec:did}.

Second, the phase argument was challenged. Here, too, our analysis adds significant information. We study the phase argument in more detail in Sec.~\ref{ssec:t}. 

\subsection{Direct and indirect muons}\label{ssec:did}

Without referring to the precise nature of muon-induced events (in particular, whether they involve secondary neutrons or not), there are two basic ways by which muons could induce the DAMA signal: directly, following muon passage through the detector, or indirectly, e.g. by muon interactions in a larger volume surrounding the detector. We now show that current LVD and DAMA data suffice to discriminate between these two possibilities, making the first one highly unlikely, regardless of the detailed physics responsible for converting muon energy deposit to single hit readouts.

First, one may suppose that DAMA counts arise from muons traversing the NaI(Tl) detector material itself~\cite{Nygren:2011xu}, or producing an enhanced yield of neutrons in the 15cm-thick led shield attached to the detector~\cite{Ralston:2010bd}. The rate of direct muons in the DAMA/LIBRA detector is 
\beq\bar R_{\rm direct\,\mu}=A\times\bar I_\mu\approx10 \;{\rm muons/day},\eeq 
where we used the LVD intensity measurement $\bar I_\mu\approx28$ muons/m$^2$/day and estimated the effective area of the DAMA/LIBRA detector by $A\approx0.35$ m$^2$. For comparison, the DAMA/LIBRA count rate in the 2-4 KeV signal region is $\bar s\,M\,\Delta E\approx530$ cpd, as can be read from Eq.~(\ref{eq:damas}) using $M=$232 kg for the active scintillator mass relevant for the first five annual cycles~\cite{Bernabei:2008yi}. 

We learn that the direct muon possibility requires a large yield, $y\approx50$ signal counts/muon, to accommodate the signal rate. By itself, the requirement of a large yield does not seem particularly constraining. As noted in~\cite{Nygren:2011xu}, muons traversing the DAMA apparatus deposit some $\mathcal{O}({\rm GeV})$ energy in the detector, leaving room for small efficiencies of $\mathcal{O}\left(10^{-4}\right)$ in producing tens of KeV light pulses. The problem with direct muons is not related to efficiencies. Rather, it has to do with the fact that a small number of seed muons are taken to be responsible for a much larger number of eventual counts.

Regardless of the statistical distribution of the yield $y$, the relative spread of the induced DAMA counts in a time bin $t_i$ cannot be smaller than that of the small number of direct muon events: 
\beq\label{eq:perr}\frac{\sigma_{s_i}}{s_i}\,>\,\frac{\sigma_{R_{\rm direct\,\mu,i}}}{R_{\rm direct\,\mu,i}}\,\approx\,30\%\times t_i^{-1/2},\eeq
with $t_i$ given in days, $s_i$ the DAMA count rate per unit mass per unit energy and $\sigma_{s_i}$ the error bar on the rate (assumed to be statistics dominated). For a typical integration time of order 50 days, Eq.~(\ref{eq:perr}) predicts a spread of at least four percent in DAMA/LIBRA residuals, a factor $\gsim5$ larger than the spread visible in practice. 
We conclude that the direct muon scenario is  unlikely. This scenario would predict a much larger spread to the DAMA residuals than that observed~\footnote{Eq.~(\ref{eq:perr}) assumes that all of the $y\sim50$ counts arising from a direct muon hit occur in the time bin $t_i$ containing the hit. This reasoning is modified if the counts exhibit very long delay $t_{\rm delay}\gg t_i$ after the muon passage, in which case we would need to replace $t_i\to t_{\rm delay}$ in Eq.~(\ref{eq:perr}). However, to remedy the discrepancy in direct muon statistics, $t_{\rm delay}\sim20\,t_i>1$ year is required, which (a) seems unrealistically long and (b) implies a very extended rise time to the count rate, absent in the data.}.

The second possibility is that DAMA events are induced by muon interactions in a large volume around the detector~\cite{Ralston:2010bd,Kudryavtsev:2008zz,wul}. In this case, the yield $y$ can be $\mathcal{O}(1)$. For given yield $y$ we can estimate the effective area through which muons should be collected,
\beq\label{eq:Aeff} A_{\rm eff}=\frac{\rm DAMA/LIBRA\,counts/day}{y\,\times\,\rm muons/m^2/day}=\frac{\bar s\,M\,\Delta E}{y\,\bar I_\mu}\approx\frac{20}{y}\,\rm m^2.\eeq
In Eq.~(\ref{eq:Aeff}) we took muons to cause the signal in the 2-4 KeV energy range with $\Delta E=2$ KeV, where DAMA observe most of the modulation. If muons are responsible to the signal in a larger energy range, then $A_{\rm eff}$ should grow proportionally. 

Note that the mean single hit rate reported by DAMA is essentially flat as a function of energy at least up to 10 KeV, except for a shallow bump of order 10 percent of the baseline which occurs between 2-4 KeV~\footnote{The location of the 2-4 KeV bump agrees with expectations from internal $^{40}K$ contamination~\cite{Kudryavtsev:2009gd}.}. As pointed out in~\cite{Nygren:2011xu}, demanding the muon signal to diminish below 10 KeV implies that some other (unmodulated) background must grow to keep the observed spectrum flat. As also pointed out in~\cite{Nygren:2011xu}, this curious fine tuning problem pertains to the dark matter hypothesis just as well as it does to background-related alternatives.

We can use the DAMA data to constrain the yield $y$. To do this, we compute the quantity
\beq\label{eq:q} q_i=\frac{\sigma_{s_i}^2}{s_i}\,M_i\,\Delta E\,\epsilon_i\,t_i,\eeq
where $M_i$ is the exposed mass and $\epsilon_i$ is the duty cycle during the time bin $t_i$. 
If the DAMA signal originates from a seed Poisson process with average yield $y$, then one expects $\langle q_i\rangle\approx y$. 
Using the 2-4 KeV DAMA/LIBRA1-5 residuals data with $M_i=232$ kg and $\epsilon_i=60\%$, we obtain $\langle q\rangle\sim1.7$.  After comparing the distribution of $q_i$ obtained from the true data to that from several sets of cosine-modulated Poisson random processes, we conclude that any background-related explanation for the DAMA signal should exhibit $y\lsim3$.

Using Eq.~(\ref{eq:Aeff}) with $y=1.7$, in agreement with the scaling of DAMA/LIBRA error bars, gives $A_{\rm eff}\sim12$ m$^2$. This could imply that the relevant muons interact within a volume extending roughly 1.5-2 meters to either side from the center of the detector, the dimensions of the DAMA/LIBRA apparatus including the surrounding meter-thick concrete housing~\cite{Bernabei:2008yh}.

\subsection{Temporal correlation}\label{ssec:t}

Ref.~\cite{Ralston:2010bd} compared the DAMA signal with ICARUS measurements of the neutron flux at LNGS~\cite{icarus}, finding a troubling correlation. We show the ICARUS data in Fig.~\ref{fig:1} (red marks). The ICARUS data set contains five measurements, extending from April 1997 to May 1998, in which  neutron events are associated with proton recoils above 3.5 MeV in a home-made organic liquid scintillation detector. Clear modulation of order few percent is present. Ref.~\cite{Ralston:2010bd} fitted a sinusoidal curve to the ICARUS data, discovering that the same fit interpolates nicely the DAMA residuals. 
While certainly alarming, the correlation reported in~\cite{Ralston:2010bd} is based on extrapolation of a loosely constrained fit. 

Ref.~\cite{Nygren:2011xu} argued against the statistical inference employed by the DAMA collaboration in~\cite{Bernabei:2009du} to reject the muon hypothesis, making the following straightforward observations:
\begin{itemize}
\item In quoting a $>5\sigma$ discrepancy, DAMA ignored the reported error on the LVD muon fit. In fact, the reported LVD phase is July 5$\pm$15 days. The combined DAMA/LIBRA and DAMA/NaI phase, May 26$\pm$7, is thus less than $3\sigma$ away from the muon phase.
\item DAMA assumed as prior a one year period to their fit. Adding the period to the fit would further reduce the statistical discrepancy.
\item Importantly, Ref.~\cite{Nygren:2011xu} observe that the fine-sampled LVD muon data does not follow a true sinusoidal behavior. If the underlying process for DAMA is also non-sinusoidal, fits to the DAMA data in which a sinusoidal form was assumed as prior will be affected by systematics, potentially invalidating statistical inference.
\end{itemize}

\begin{figure}[h]\begin{center}
\includegraphics[width=15cm]{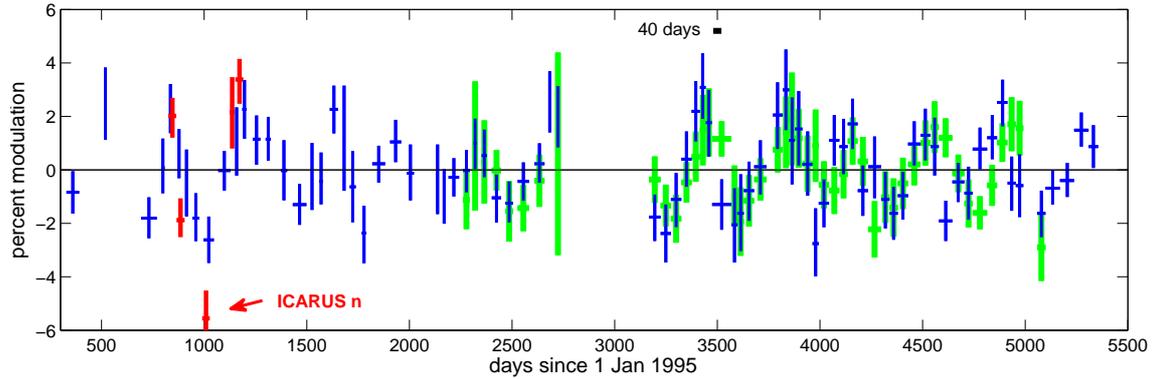}
\caption{DAMA residuals (blue) and binned muon intensity residuals (green). 
ICARUS neutron measurements during 1997-1998 are added (red).}
\label{fig:2}
\end{center}
\end{figure}

To explore the muon hypothesis in more detail, in Fig.~\ref{fig:2} we bin the muon residuals on the same binning as the DAMA residuals. 
To emphasize the time scale, a 40 day period -- equal to the difference between LVD's and DAMA's sinusoidal best fit phases -- is depicted in Fig.~\ref{fig:2} as a black line above day 3500.

To estimate the significance of a phase discrepancy between the DAMA and muon data we proceed as follows. 
We define the muon hypothesis as saying that the DAMA count rate in time bin $t_i$ is given by
\beq\label{eq:test0} s_i=\frac{y\,N_{\mu,i}}{M_i\,\Delta E\,\epsilon_i\,t_i},\eeq
where $N_{\mu,i}$ is Poisson distributed with mean
\beq\label{eq:test}\left\langle N_{\mu,i}\right\rangle=A_{\rm eff}\,I_{\mu,i}\,\epsilon_i\,t_i,\eeq
with $A_{\rm eff}$ taken from Eq.~(\ref{eq:Aeff}) and  $\epsilon_i=60\%$. In Eq.~(\ref{eq:test}), $I_{\mu,i}$ is the muon intensity in time bin $t_i$.
For simplicity, we adopt the yield $y=2$ and neglect sources of background other than those correlated with muons. We generate multiple realizations of $N_{\mu,i}$ for the time bins of DAMA/LIBRA1-5. For each realization we define and remove the annual baselines, and then fit the resulting time sequence to a cosine function, mimicking the DAMA procedure. We thus obtain distributions of the following quantities: (1) the best fit phase $t_0$, allowing both the amplitude and the period to float; (2) the best fit $\chi^2$, minimizing over the amplitude with period and phase fixed to 1 year and day 152.5, respectively. 

\begin{figure}[h]\begin{center}
\includegraphics[width=7cm]{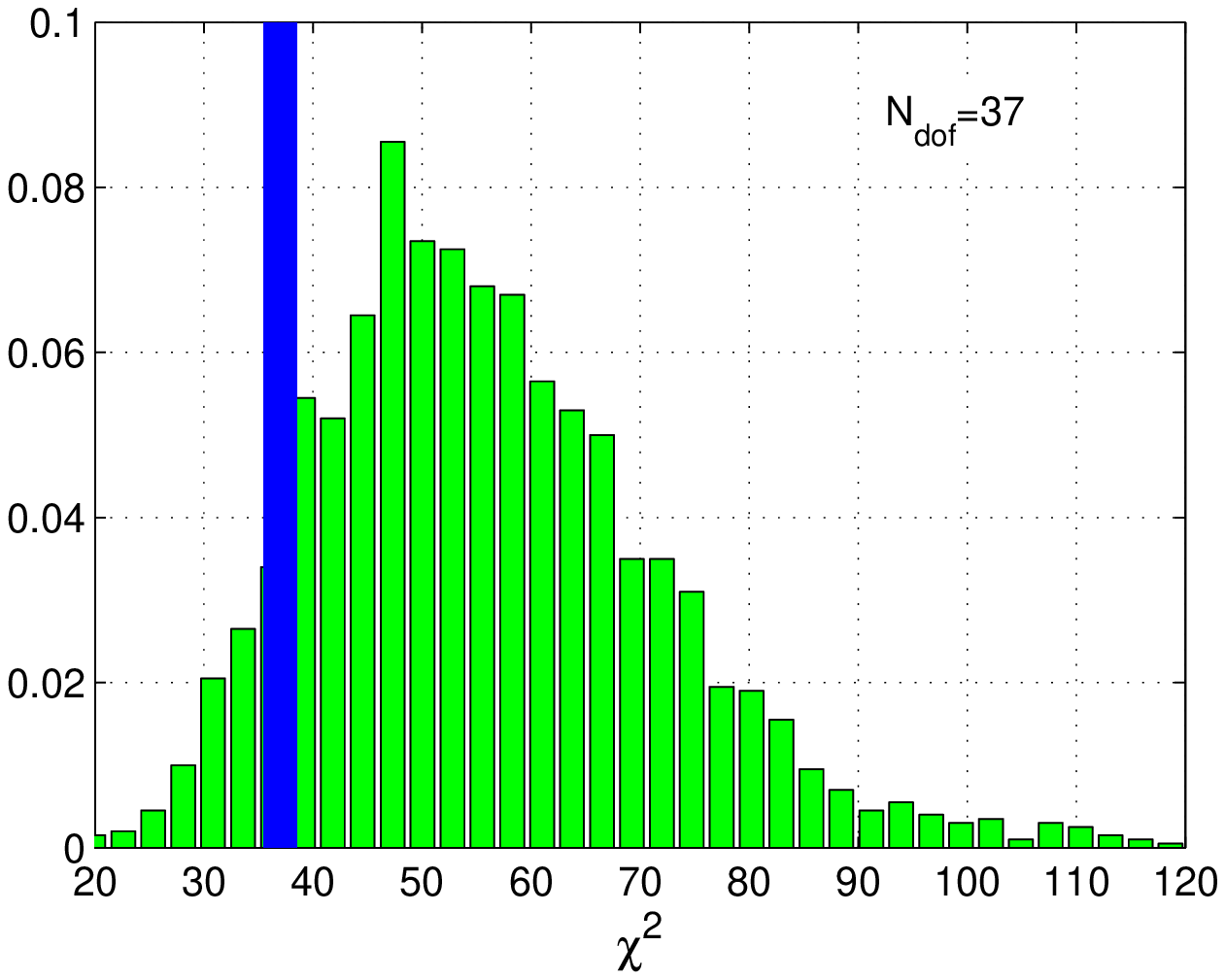}\hfill
\includegraphics[width=7cm]{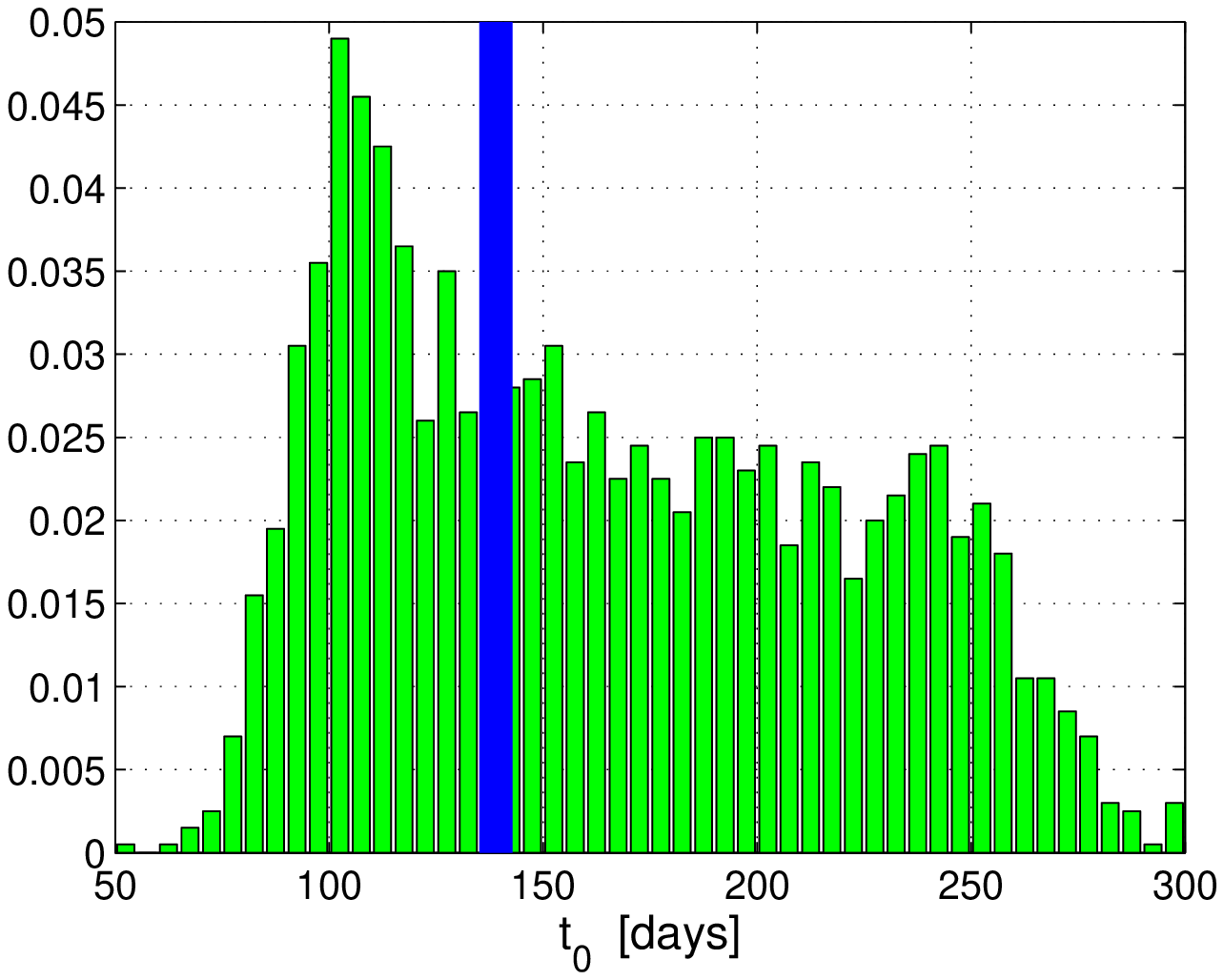}
\caption{Distributions of the best fit $\chi^2$ (left panel) and phase $t_0$ (right panel), generated for multiple realizations of mock data under the muon hypothesis. DAMA fit results are shown as blue vertical bars.}
\label{fig:test}
\end{center}
\end{figure}

The resulting distributions of $\chi^2$ and $t_0$ are shown in the left and right panels of Fig.~\ref{fig:test} (green). The number of dof in the $\chi^2$ fit are 37, with 38 DAMA/LIBRA1-5 time bins and one fitted amplitude parameter. What interests us here is the distribution of $\chi^2$ and $t_0$ for the mock data, in comparison with the results $\chi^2/dof\approx1$ and $t_0\approx140$ days~\cite{Bernabei:2010mq}, obtained by DAMA for their true data and shown as blue bands in Fig.~\ref{fig:test}. We conclude that the simplistic muon hypothesis defined by Eqs.~(\ref{eq:test0}-\ref{eq:test}) is consistent with the results reported by DAMA.

\subsection{The usefulness of annual baseline data}\label{ssec:qtest}

If DAMA counts arise from muons -- be it through secondary particles produced by interactions in the surrounding volume, activation of impurities or even in some mundane manner via disturbances to the electronics -- then the yearly baseline count rates should reflect long term trends in the muon data. In the energy range 2-4 KeV, these annual baseline rates should be measured to accuracy of $\sim0.25\%$. 

\begin{figure}[h]\begin{center}
\includegraphics[width=15cm]{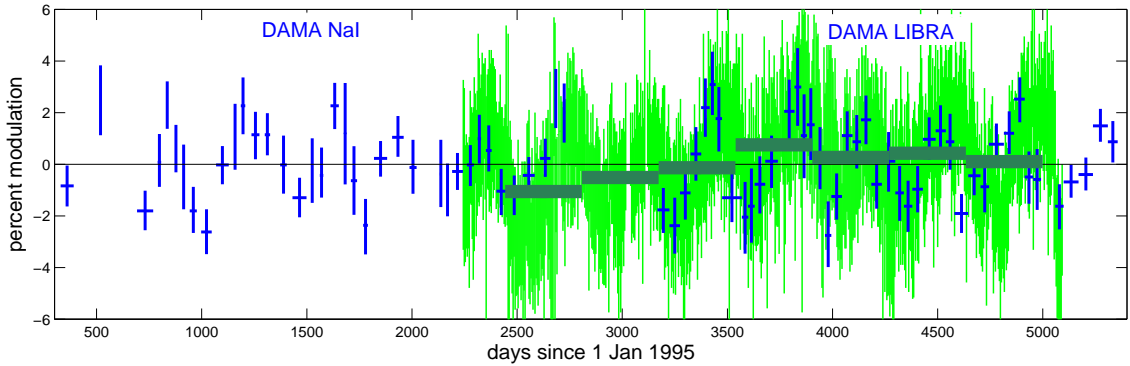}
\caption{Long term modulation in the annually-averaged muon rate (thick dark green) with respect to the seven year average in the period Sep 9 2001 until Sep 9 2008. The thickness of the horizontal bars is $\pm0.25\%$, representing the expected DAMA resolution for the energy range 2-4 KeV. Here, in contrast to Fig.~\ref{fig:1}, the short term muon intensity residuals (green) were defined with respect to the seven year baseline rather than to the different annual baselines.}
\label{fig:mu_baseline}
\end{center}
\end{figure}

In Fig.~\ref{fig:mu_baseline} we plot again the DAMA and muon data. This time, instead of removing the baselines for each year separately, we show the muon residuals defined with respect to the seven year average in the period 9 Sep 2001 until 9 Sep 2008. The muon intensity exhibits long term modulation with period $\sim6$ years and non negligible amplitude $\gsim1\%$. The annual baselines of the muon intensity are shown as dark green bands. The width of each band is $\pm0.25\%$, the estimated DAMA/LIBRA resolution for a one year integration time. 

\section{Conclusions}\label{s:disc}

The DAMA anomaly has been in existence for over a decade. No solution in terms of a background process has thus far been verified. 
In this paper we compared the DAMA signal to the underground muon flux at LNGS. We found that the two observables are tightly correlated. The correlation persists throughout the 13 annual cycles of operation of DAMA/LIBRA and its former DAMA/NaI (see Figs.~\ref{fig:1} and~\ref{fig:DAMA_T_PDM}), motivating muon-induced events as an explanation for the DAMA result. 

Our work adds to Refs.~\cite{Ralston:2010bd,Nygren:2011xu}, which suggested various mechanisms by which muon-induced interactions could potentially bypass the DAMA background rejection criteria, and pointed out that both the amplitude and the phase of the annual modulation in the underground flux of neutrons and muons at LNGS are not inconsistent with DAMA's.

The underground muon flux exhibits short and long term variations of order percent on time scales ranging from weeks to several years, deviating systematically from pure sinusoidal form. This behavior is not expected from the dark matter signal. The DAMA collaboration can potentially exclude a muon-induced background by reporting their high statistics annual baseline count rates in the signal region. 

While there are suggestions, the actual mechanism by which muons may eventually lead to the DAMA signal remains unknown. Any hints which may hide in the data are thus important.
Inspecting the spread of DAMA residuals, we showed that it is unlikely for the DAMA signal to arise solely from muons traversing the detector material itself. This scenario, suggested in~\cite{Nygren:2011xu}, would lead to a large irreducible statistical spread in DAMA events, which is not manifest in the data. In general, the same argument disfavors high yield processes in which few seed interactions deliver a large amount of energy into the DAMA detector material, subsequently transformed into multiple KeV light pulses in the signal region. Using available DAMA data we showed that the yield of such process must be moderate, $y\lsim3$. The signal may arise from muon interactions throughout a volume extending $\sim2-3$ meters to either side of the active detector.

Lastly, when considering the implications of Fig.~\ref{fig:1}, a comment is in order. One should bear in mind the accidental phase proximity of the expected dark matter peak (early June) and solar solstice (late June, around which time the atmosphere in the northern hemisphere is warmest leading to a maximum in the underground muon intensity). This accidental coincidence should be regarded as a prior, making it harder to discriminate a muon-induced background from the dark matter hypothesis~\cite{Freese:1987wu}, but not adding to the relative significance of the former. Having made this cautionary comment, it is still remarkable that the modulation amplitudes of the DAMA signal and the muon intensity are so close.

\begin{appendix}
\section{The muon flux -- temperature correlation}\label{a1}

High energy muons are generated through meson decays in the stratosphere and penetrate deep underground.
Annual temperature changes induce a corresponding modulation in the muon rate,
\beq\label{RT}\frac{\Delta R_\mu}{R_\mu}=\alpha_T\frac{\Delta \Te}{\Te},\eeq
where $\alpha_T$ is a proportionality coefficient which depends on the cosmic ray composition in the upper levels of the atmosphere, and $\Te$ is an effective temperature~\cite{gaisser,Ambrosio:1997tc}. 
Here we provide a quick analytical derivation of Eq.~(\ref{RT}) using a toy model. We note that Eq.~(\ref{RT}) is well known in the literature since early times~\cite{forro,barrett,gaisser,Ambrosio:1997tc} and so this part is mainly added for completeness. We then proceed to compute $\Te$ and use it first as independent verification of the time variability of the muon rate on different time scales and, second, to cover the entire 13 years of DAMA/NaI and DAMA/LIBRA operation.

\paragraph{Understanding the muon flux -- temperature correlation.} 

Let us compute the effect of temperature variations on the muon intensity in a toy model, approximating the atmosphere as a single layer of air with uniform density and temperature, considering an isotropic distribution of muons, and ignoring the precise cosmic ray composition and the details of muon propagation in rock. We work with energy in GeV units, $\epsilon\equiv E/$GeV.

Atmospheric muons are produced in the decay of pions and kaons, generated in turn by cosmic ray collisions with air. Here we consider only the dominant pion source. At production, pions inherit the spectral index of the primary cosmic rays such that their flux obeys
\beq J_\pi(\epsilon)\propto\epsilon^{-\gamma}\eeq
with $\gamma=2.75$.
The spectrum of pion-decay muons depends on the competition of two processes:
\begin{enumerate}
\item pion decay, controlled by the observer frame lifetime $t_\pi(\epsilon)=\frac{E\tau_\pi}{m_\pi}=1.8\cdot10^{-7}\,\epsilon$ s
\item collisions with air, with characteristic time $t_c=\frac{1}{\sigma nc}\sim10^{-5}\,\frac{1}{n}$ s, the air density $n$ given in units of $3\cdot10^{19}$ cm$^{-3}$, the relevant order of magnitude at the tens of km height where most of the muons form. The air density responds to temperature changes, $\frac{1}{n}\propto T$. Thus $t_c\propto T$.
\end{enumerate}
Equating the time scales $t_\pi$ and $t_c$ defines a single critical energy:
\beq t_\pi(\bar\epsilon)=t_c\,\Rightarrow\bar\epsilon=\frac{m_\pi t_c}{\tau_\pi}\approx100\frac{1}{n}.\eeq
At energy $\bar\epsilon$ collision losses and decay are equally important. We see that $\bar\epsilon\propto T$.

Collisions with air degrade the energy of pions, such that the differential probability of a pion produced at energy $\epsilon_{\pi,0}$ to decay at energy $\epsilon_{\pi,d}$ is:
\beq\label{eq:pi}\frac{dP(\epsilon_{\pi,0},\epsilon_{\pi,d})}{d\epsilon_{\pi,d}}=\frac{\bar\epsilon e^{-\left(\frac{\bar\epsilon}{\epsilon_{\pi,d}}-\frac{\bar\epsilon}{\epsilon_{\pi,0}}\right)}}{\epsilon_{\pi,d}^2}.\eeq
We also need the distribution of muons from pion decay:
\beq\label{eq:mu} f_\mu(\epsilon_\pi,\epsilon_\mu)=\left\{\begin{array}{ccc}\frac{1}{\epsilon_\pi(1-r_\pi)}&,&\epsilon_\mu\in[r_\pi\epsilon_\pi,\epsilon_\pi]\\0&,&{\rm o.w.}\end{array}\right\},\;\;\;r_\pi=\frac{m_\mu^2}{m_\pi^2}\approx0.57.\eeq
Using Eqs.~(\ref{eq:pi}-\ref{eq:mu}) we have
\beq\label{eq:Tmu} J_\mu(\epsilon_\mu)&=&\int_{\epsilon_\mu}^\infty d\epsilon_{\pi,d}f_\mu(\epsilon_{\pi,d},\epsilon_\mu)\int_{\epsilon_{\pi,d}}^\infty d\epsilon_{\pi,0}J_\pi(\epsilon_{\pi,0})\frac{dP(\epsilon_{\pi,0},\epsilon_{\pi,d})}{d\epsilon_{\pi,d}}
=\frac{J_\pi(\epsilon_\mu)}{1-r_\pi}\,\mathcal{F}\left(\frac{\epsilon_\mu}{\bar\epsilon}\right),\no\\
\mathcal{F}(x)&=&\frac{1}{x}\,\int_{r_\pi}^1d\eta\eta^\gamma\,\int_0^1 d\xi\xi^{\gamma-2}e^{-\frac{1}{x}\eta\left(1-\xi\right)}.
\eeq
For $x<0.1$, $\mathcal{F}(x)\approx0.3$. For $x>1$, $\mathcal{F}(x)\approx0.13/x$. Thus for $\epsilon_\mu<\bar\epsilon$, the muon flux at sea level inherits the production spectrum of the pions, while for $\epsilon_\mu>\bar\epsilon$ the muon flux is softer by one power of energy.

Computing the muon rate above some threshold $\epsilon_{th}$ gives:
\beq R\propto\int_{\epsilon_{th}}^\infty d\epsilon J_\mu(\epsilon)\propto\bar\epsilon^{1-\gamma}\,\int_{\frac{\epsilon_{th}}{\bar\epsilon}}^\infty dxx^{-\gamma}\mathcal{F}(x).\eeq
We learn that if $\epsilon_{th}>\bar\epsilon$, then $R\propto\bar\epsilon$, while if $\epsilon_{th}<\bar\epsilon$, then $R$ is approximately independent of $\bar\epsilon$. 
As shown above, $\bar\epsilon\propto T$ and thus for $\epsilon_{th}>\bar\epsilon$ we find
\beq\frac{\Delta R}{R}=\alpha_T\frac{\Delta T}{T},\eeq
with $\alpha_T\approx1$.
This completes the derivation of Eq.~(\ref{RT}), up to refined considerations which are needed in order to determine the precise value of $\alpha_T$. Note that a 3 kmwe rock overburden at LNGS is equivalent to $\epsilon_{th}\gsim10^3$ (1 TeV) for the surface muon energy, sufficiently large compared to $\bar\epsilon$ for pions.
The MACRO collaboration~\cite{Ambrosio:1997tc}  measured the value of $\alpha_T$ at their apparatus in LNGS, obtaining $\alpha_T=0.98\pm0.12$ for their full data set in the years 1991-1994 and $\alpha_T=0.83\pm0.13$ for a partial data set extending over 1993-1994.

\paragraph{Computation of $\Te$.}

The effective temperature $\Te$ corresponds to an average of the temperature at different layers in the atmosphere, weighted by the probability for the observed muons to be formed at each layer.
A useful approximate relation is given by~\cite{Ambrosio:1997tc}
\beq\label{Te}\Te&=&\frac{\int\frac{dX}{X}T(X)\left(e^{-X/\Lambda_\pi}-e^{-X/\Lambda_N}\right)}{\int\frac{dX}{X}\left(e^{-X/\Lambda_\pi}-e^{-X/\Lambda_N}\right)},
\eeq
where $\Lambda_\pi=160$ gcm$^{-2}$, $\Lambda_N=120$ gcm$^{-2}$ and the integration is over the slant depth $X$ (in units of gcm$^{-2}$), for which we use an approximate isothermal expression~\cite{gaisser}
\beq X(h)\approx X_0e^{-h/h_0},\;\;X_0=1030\,{\rm gcm^{-2}},\;h_0=6.4\,{\rm km}\eeq
with $h$ the geophysical height in km.

We obtain atmospheric temperature records from the Integrated Global Radiosonde Archive (IGRA) of the National Climatic Data Center (NCDC), US Department of Commerce~\cite{NCDC}. Archival information is available for the Pratica di Mare station in Italy, located at a distance of 124 km from LNGS. IGRA data contains temperature measurements at different discrete geophysical heights, typically derived up to few times per day. Using these data we compute $\Te$, approximating the integrals in Eq.~(\ref{Te}) by discrete sums. 

\begin{figure}[h]\begin{center}
\includegraphics[width=14cm]{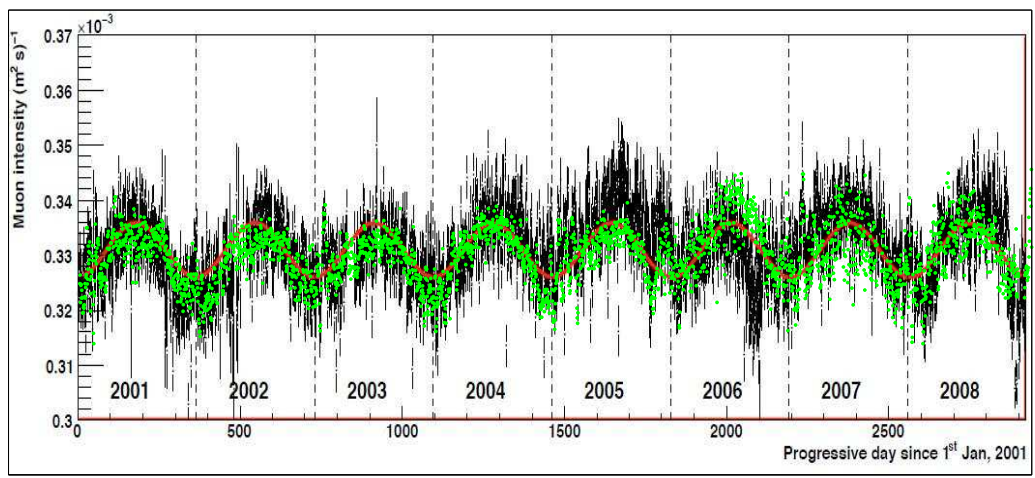}
\caption{Muon flux at LVD (black data, red sinusoidal fit, taken from~\cite{lvd}) and $\alpha_T\left(\Delta T/T\right)$ at the stratosphere (green).}
\label{fig:lvd}
\end{center}
\end{figure}

In Fig.~\ref{fig:lvd} we plot the temperature modulation above the Pratica di Mare station (green) together with the LVD muon rate data (black markers with red sine fit, adopted from~\cite{lvd}). The correlation is clear. Occasional systematic deviations can be traced to the geographical distance between the weather station and LNGS. 

\begin{figure}[h]\begin{center}
\includegraphics[width=15cm]{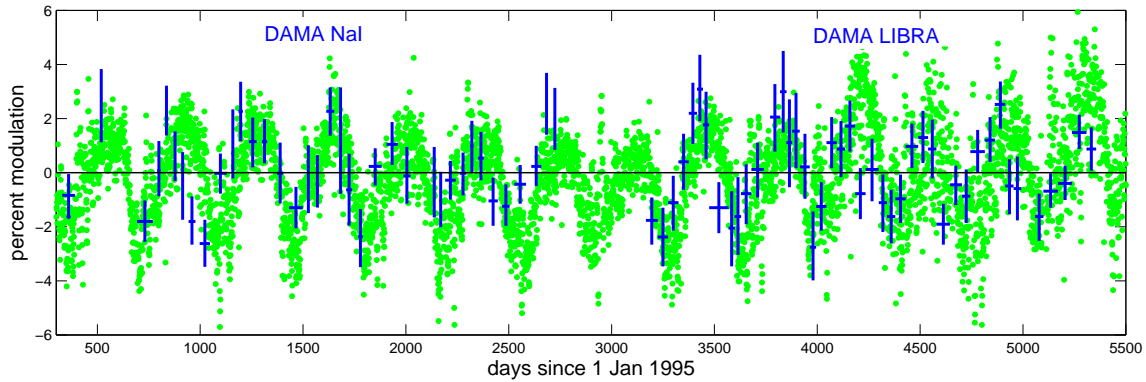}
\caption{DAMA residuals (blue) and stratospheric temperature residuals $\Delta T_{eff}/T_{eff}$ (green), in percent from the respective baselines.}
\label{fig:DAMA_T_PDM}
\end{center}
\end{figure}

Lastly, in Fig.~\ref{fig:DAMA_T_PDM} we plot the DAMA residuals together with the daily stratospheric temperature modulation. The temperature analysis allows us to follow the 13 years of DAMA data taking, extending beyond the 8 years of LVD muon records.

\end{appendix}

\section*{Acknowledgments}

We thank M.~Hass, B.~Katz, M.~Milgrom, D.~Nygren, S.~Vaintraub, T.~Volansky, B.~Ziv and especially T.~Montaruli, Y.~Nir and E.~Waxman for useful  discussions. We are grateful to R.~Bernabei for clarifications regarding the DAMA analysis.


\end{document}